\documentclass[a4paper,12pt]{article}
\usepackage[margin=2cm]{geometry}
\usepackage{comment}
\usepackage{amssymb,extarrows,graphicx,subfigure,setspace}
\usepackage{cite}
\usepackage{slashed}
\usepackage{tensor}
\usepackage[toc,page]{appendix}
\usepackage{color}
\usepackage{physics}
\usepackage{hyperref}
\usepackage{dirtytalk}
\hypersetup{colorlinks=true, linkcolor=blue, citecolor=red, linktoc=page}
\makeatother

\usepackage{amsmath}
\newmuskip\pFqmuskip

\newcommand*\pFq[6][8]{%
  \begingroup 
  \pFqmuskip=#1mu\relax
  \mathcode`\,=\string"8000
  \begingroup\lccode`\~=`\,
  \lowercase{\endgroup\let~}\pFqcomma
  {}_{#2}F_{#3}{\left[\genfrac..{0pt}{}{#4}{#5};#6\right]}%
  \endgroup
}
\newcommand{\pFqcomma}{\mskip\pFqmuskip}

\usepackage{mathrsfs}  
\usepackage{hyperref}

\newcommand{\be}{\begin{equation}}
\newcommand{\bea}{\begin{eqnarray}}
\newcommand{\eea}{\end{eqnarray}}
\newcommand{\ba}{\begin{array}}
\newcommand{\ea}{\end{array}}
\newcommand{\ee}{\end{equation}}
\newcommand{\bes}{\begin{equation*}}
\newcommand{\beas}{\begin{eqnarray*}}
\newcommand{\eeas}{\end{eqnarray*}}
\newcommand{\bas}{\begin{array*}}
\newcommand{\eas}{\end{array*}}
\newcommand{\ees}{\end{equation*}}

\numberwithin{equation}{section}
\begin{document}
	\onehalfspacing
	\noindent
	
	\begin{titlepage}
		

		\vspace*{20mm}
		\begin{center}
			
			{\Large {\bf Thermodynamic Extremality in Power-law AdS  \\[0.2cm] Black Holes: A Universal Perspective }
			}
			
			\vspace*{15mm}
			\vspace*{1mm}
		{\bf \large Ankit Anand }
		\footnote{ E-mail : ankitanandp94@gmail.com}
		\vskip 0.5cm
		{\it Department of Physics, Indian Institute of Technology Kanpur, Kanpur 208016, India.}
			\vspace{0.2cm}

			\vspace*{1cm}
		\end{center}
		
\begin{abstract}

This study investigates the universal relation between Goon and Penco (GP) proposed within the frameworks of Power-Maxwell, Power-Yang-Mills, and Maxwell-Power-Yang-Mills black holes. We begin by analyzing these black holes' thermodynamics and then calculating the perturbed metric and thermodynamic quantities by perturbing the action. Our objective is to examine the consistency of the GP relation across various power-law terms in the field equations, aiming to gain deeper insights into the nature of these black holes. The GP connection remains robust across different power spacetimes, indicating that this relation is a universal feature of black holes.

\hspace{5 cm}\\
\hspace{5 cm}\\
\hspace{5 cm}\\
\hspace{5 cm}\\
\hspace{5 cm}\\
\hspace{5 cm}\\
\hspace{5 cm}

\end{abstract}
\end{titlepage}

\newpage
\tableofcontents


\section{Introduction}
\label{Sec: Introduction}

In general relativity, black holes are compact objects that represent the ultimate state of massive stellar collapse. They are characterized by a spherical boundary known as the event horizon, beyond which no outgoing geodesics exist. Any geodesic entering this horizon inevitably progresses toward the singularity at the center of the black hole, where conventional physical laws break down \cite{Penrose:1964wq}. Consequently, no radiation can escape from within a black hole. From a classical perspective, the mass of a black hole can be decomposed into reducible and irreducible components. The reducible energy can be modified through interactions, while the irreducible mass remains constant \cite{Christodoulou:1970wf, Christodoulou:1971pcn}. Hawking's introduction of black hole radiation, known as Hawking radiation, recast black holes as thermal entities with a temperature proportional to their surface gravity \cite{Hawking:1974rv, Hawking:1975vcx}. This led to the development of the concept of black hole entropy by Bekenstein \cite{Bekenstein:1973ur, Bekenstein:1974ax}, which is proportional to the surface area of the black hole and is referred to as Hawking–Bekenstein entropy. The foundational laws of black hole thermodynamics were formulated in the 1970s \cite{Bardeen:1973gs}, including the zeroth, first, second, and third laws, which are based on principles such as surface gravity, the fundamental form of the first law, and the non-decreasing nature of a black hole's area. A mass formula for Kerr–Newman black holes was established, which relates the mass to the black hole’s area and bound \cite{Smarr:1972kt}. These advances redefined black holes as thermally radiating objects governed by thermodynamic principles and laws.

In the domain of quantum gravity, Vafa introduced the Weak Gravity Conjecture to address the charge-to-mass ratio \cite{Vafa:2005ui, Arkani-Hamed:2006emk}. The WGC asserts that any gauge force should exert a greater effect than gravity, formulated as $\frac{Q}{M} \geq 1$, with equality achieved in extremal black holes. This conjecture, grounded in quantum gravity principles, is particularly significant in the absence of global symmetries. In quantum gravity, evaporating black holes that do not emit charged particles via Hawking radiation release particles independent of their global charges \cite{Banks:2010zn, Harlow:2022ich}. The WGC is supported by various studies that align with this conjecture, including research introducing correction terms to prevent naked singularities, as the absence of naked singularities supports the WGC inequality \cite{Cheung:2018cwt}.

Building upon WGC studies, Goon and Penco explored the universality of the thermodynamic relationship between entropy and extremality under perturbations \cite{Goon:2019faz}. They derived a relation between the derivative of mass and entropy as 
\begin{equation}\label{Goon Penco relation}
    \frac{\partial M_{ext}(\Vec{\mathcal{Q}},\epsilon)}{\partial\epsilon} = \lim_{M \to M_{ext}(\Vec{\mathcal{Q}},\epsilon)} -T \left( \frac{\partial S(M,\Vec{\mathcal{Q}},\epsilon)}{\partial\epsilon} \right)_{M,\Vec{\mathcal{Q}}} \ ,
\end{equation}
where \( M \), \( \Vec{\mathcal{Q}} \), and \( \epsilon \) denote mass, additional quantities, and the perturbative parameter, respectively. Perturbations in free energy establish a relationship among mass, temperature, and entropy with corrections. The leading-order expansion of these perturbative parameters reveals an approximate relation linked to higher-derivative corrections \cite{Wang:2022sbp}, connecting shifts in entropy to the charge-to-mass ratio \cite{Cheung:2018cwt, Cheung:2019cwi, Kats:2006xp}. The Vafa conjecture\cite{Vafa:2005ui}, central to the Swampland program, delineates effective field theories consistent with quantum gravity from those not. Within this framework, the Weak Gravity Conjecture emerges as a fundamental criterion, asserting that gravity must be the weakest force. The WGC prevents the existence of stable extremal black holes by requiring the presence of super-extremal particles, ensuring decay via charged emission. This condition aligns with the Swampland Conjecture \cite{Alipour:2024eoi, NooriGashti:2024ldh, Alipour:2024jgn, Anand:2024hpj, Sadeghi:2024dnw, NooriGashti:2024gnc, Sadeghi:2024ixf, Sadeghi:2023zdc} and constraints on de Sitter vacua. Thus, the WGC serves as a specific realization of Vafa’s Swampland conjecture, imposing critical limits on effective field theories. This Goon and Penco relation mainly supports the WGC when the mass shift in extremal black holes is proportional to the entropy shift with a negative constant \cite{Reall:2019sah}, as examined in \cite{Ma:2023qqj}. Based on Goon and Penco's relation, significant progress has been made in analyzing various AdS spacetimes, such as charged BTZ black holes and Kerr–AdS black holes, from the WGC perspective \cite{Cano:2019ycn, Cremonini:2019wdk, Cano:2019oma, Sadeghi:2020xtc, Wei:2020bgk, Chen:2020rov, Chen:2020bvv, Sadeghi:2020ciy, Ma:2020xwi, McPeak:2021tvu, McInnes:2021frb, Etheredge:2022rfl, Sadeghi:2022xcr, SADEGHI2020100626, Sadeghi:2020ciy, Sadeghi:2020ntn}.

General relativity \cite{Volkov:1989fi, Brihaye:2006xc, Mazharimousavi:2008ap, Bostani:2009zf, Gao:2003ys, Devecioglu:2014iia, Bellucci:2011gz} and higher-order derivative gravities \cite{HabibMazharimousavi:2008zz, HabibMazharimousavi:2008ib, Mazharimousavi:2009mb} have explored black holes that include two gauge fields (the Maxwell field and the Yang-Mills field) coupled through gravity. From a physics perspective, the Yang-Mills field is limited to acting inside nuclei, whereas electromagnetism has long-range effects and prevails beyond the nucleus of natural matter. Black holes in the Einstein-power-Maxwell, Einstein-power-Yang-Mills, and Einstein-Maxwell-power-Yang-Mills theories of gravity are the focus of this work.

The study of gravitational theories considering nonlinearity in the Maxwell and Yang-Mills fields has attracted much attention lately. Because they are nonsingular, black hole solutions in nonlinear electrodynamics are highly intriguing \cite{Ayon}. The Born-Infeld electrodynamics \cite{Born}, which smoothed out divergences at the origin caused by the linear electric field, is particularly interesting in this context. In power-invariant theory, a class of black hole solutions is found with Lagrangian density denoted by $(F_{\mu\nu} F^{\mu\nu})^\gamma$, where $\gamma$ is an arbitrary rational number \cite{Mae}. Numerous researchers have investigated alternative non-linear models in which the non-abelian Yang-Mills field is linked to gravity in general relativity, following their investigation of Einstein-PMI gravity's black hole solutions. In \cite{Mazharimousavi:2009mb}, the authors studied potential black hole solutions supplied by the power of the Yang-Mills $(YM)$ invariant as $(F_{\mu\nu}^{(a)} F^{(a)\mu\nu})^\xi$. Setting $\xi = 1$ recovers the $4$-dimensional Einstein-Yang-Mills $(EYM)$ black holes in $AdS$ spacetime \cite{Hal12, Hal22, Hendi, Zhan}. The Van der Waals-like phase transition and critical behavior in the extended thermodynamics of $AdS$ black holes in Einstein-power-Maxwell and Einstein-power-Yang-Mills theories were examined in \cite{Chandra}.

In this study, we examine the universal relation posited by Goon and Penco within the context of power-law Maxwell and power-law Yang-Mills black holes, a domain that has not been thoroughly explored in these spacetimes. The paper is structured as follows: In the next section \ref{Sec:Universal Law and Power law Black holes}, we present the solution for power Maxwell and power-Yang-Mills black holes. We explore their thermodynamic properties, highlighting key characteristics and behaviors that arise from this specific gravitational context. In Section \ref{Sec: Extrmality Relations}, we introduce a perturbative correction to the action, which allows us to derive the extremality relations that connect mass, pressure, entropy, and charges. We also examine the parameters associated with these relationships, thoroughly analyzing how these quantities interact under the influence of the perturbative correction. Finally, Section \ref{Sec: Discussion} is dedicated to a comprehensive discussion of our findings obtained throughout the paper and drawing conclusions based on our analysis.

\section{Universal Law and Power law Black holes}\label{Sec:Universal Law and Power law Black holes}

The correction term in the action 
\begin{equation} \label{Delta I}
    \Delta \mathcal{I} = \mathcal{K}  \int d^4 x \; \sqrt{-g } \;  \epsilon \; \Lambda \ ,
\end{equation}
where, $\mathcal{K}$ is some constant, $\epsilon$ is perturbation parameter and $\Lambda$ is the cosmological constant. We assume that $\epsilon$ is a small parameter, and as $\epsilon$ approaches zero, the action reduces to its uncorrected form. The total action can be written as 
\begin{equation*}
    \mathcal{I}_{\text{Total}} = \mathcal{I}\; + \; \Delta \; \mathcal{I} \ .
\end{equation*}
The metric is also modified by the changes in the theory, where $g_{\mu\nu} = \widehat{g}_{\mu\nu} \,  + \, \epsilon \, \Delta \, g_{\mu\nu} =  \widehat{g}_{\mu\nu} \,  + \, \epsilon \, \mathrm{h}_{\mu \nu}$, with \(\widehat{g}_{\mu\nu}\) being a solution of \(\mathcal{I}\) and $\mathrm{h}_{\mu \nu}$ is the perturbed metric by adding the above perturbation. 

\quad In Einstein gravity, the Smarr relation in \cite{Smarr:1973} showed that the action \(\mathcal{I}\) can be expressed in terms of the Gibbs free energy \(G(T,\Vec{\mu})\)\footnote{Here we have considered a thermodynamic system defined by its entropy \(S\) and a set of additional extensive variables \(\vec{\mathcal{Q}}\), such that the energy of the system is given by \(M(S, \vec{\mathcal{Q}})\). The first law of thermodynamics in this context takes the form:
\[
dM = TdS + \vec{\mu} \cdot d\vec{\mathcal{Q}} \ ,
\]
Where \(\mu\) represents a set of generalized chemical potentials. In the context of black holes, the variables \(Q\) may include conserved quantities such as angular momenta and \(U(1)\) charges, as well as terms not associated with any conservation law, such as the volume of the black hole. The Gibb's free energy is 
\[
G(T,\vec{\mu}) = M (T,\vec{\mu}) - TS - \vec{\mu} \cdot \vec{\mathcal{Q}} \ .
\] }. 
Perturbative corrections to the free energy \( G(T,\Vec{\mu}) \) are proportional corrections to the system's action \( I \). While boundary-term contributions may arise, they are typically negligible. This correction specifically affects the cosmological term in the action, leading to changes in the horizon radius, mass, temperature, and entropy. So, adding the correction term in actual Gibbs free energy, the total Gibbs free energy also gets corrected and can be written in the terms of corrected Gibbs free energy $\Delta G$ as
\begin{eqnarray}\label{Free Energy}
    G(T,\Vec{\mu}) \rightarrow G(T,\Vec{\mu}) + \epsilon\, \Delta \, G(T,\Vec{\mu}) \ .
\end{eqnarray}
We can compute the other quantities in terms of $G$ as
\begin{eqnarray}\label{Thermo Relations}
    S(T,\vec{\mu},\epsilon) = -\left (\frac{\partial G}{\partial T}\right )_{\vec{\mu},\epsilon}  \;\;\;\;\;;\;\;\;\;\; Q_{i}(T,\vec{\mu},\epsilon) = -\left (\frac{\partial G}{\partial \mu_{i}}\right )_{T,\mu_{j\neq i},\epsilon} \ .
\end{eqnarray}
A universal relation can be derived from \eqref{Free Energy} by postulating that \eqref{Thermo Relations} is invertible, allowing us to freely exchange the variables  \(T\), \(\vec{\mu}\), \(M\), and \(\vec{\mathcal{Q}}\). This assumption grants the flexibility to reframe the thermodynamic variables, enabling a deeper insight into their interconnected roles within the system. Additionally, we assume that the perturbative corrections to the entropy satisfy the condition:
\begin{equation}\label{Mext}
    \lim_{T\to 0} \; T \left (\frac{\partial S(T,\vec{\mathcal{Q}},\epsilon)}{\partial \epsilon}\right )_{T,\vec{\mathcal{Q}}} = 0 \ ,
\end{equation}
which can be interpreted as a manifestation of the third law of thermodynamics.

\quad We begin by examining the perturbed extremality bound, expressed as \(M > M_{\rm ext}(\vec{\mathcal{Q}},\epsilon)\), where
\[
M_{\rm ext}(\vec{\mathcal{Q}},\epsilon) \equiv \lim_{T\to 0} \, M(T,\vec{\mathcal{Q}},\epsilon) \ .
\]
To understand the impact of perturbative corrections, we compute the \(\epsilon\)-derivative of \(M\) while keeping \(T\) and \(\vec{\mathcal{Q}}\) constant. By applying \(\eqref{Thermo Relations}\) and utilizing the chain rule, we obtain:
\begin{eqnarray}
   \left(\frac{\partial M}{\partial \epsilon}\right)_{T,\vec{\mathcal{Q}}} &=& \left(\frac{\partial}{\partial \epsilon} \left(G + TS + \vec{\mu} \cdot \vec{\mathcal{Q}}\right)\right)_{T,\vec{\mathcal{Q}}} \nonumber  \\
 &=& \left(\frac{\partial G}{\partial \vec{\mu}}\right)_{T,\epsilon} \cdot \left(\frac{\partial \vec{\mu}}{\partial \epsilon}\right)_{T,\vec{\mathcal{Q}}} + \left(\frac{\partial G}{\partial \epsilon}\right)_{T,\vec{\mu}} + T \left(\frac{\partial S}{\partial \epsilon}\right)_{T,\vec{\mathcal{Q}}} + \vec{\mathcal{Q}} \cdot \left(\frac{\partial \vec{\mu}}{\partial \epsilon}\right)_{T,\vec{\mathcal{Q}}}  \nonumber \\
 &=& \left(\frac{\partial G}{\partial \epsilon}\right)_{T,\vec{\mu}}  + T \left(\frac{\partial S}{\partial \epsilon}\right)_{T,\vec{\mathcal{Q}}} \ . \nonumber
 \end{eqnarray}
where \(G\) is treated as a function of \((T,\vec{\mu},\epsilon)\), while \(S\) and \(\vec{\mu}\) are considered functions of \((T,\vec{\mathcal{Q}},\epsilon)\). The first and last terms cancel due to \(\eqref{Thermo Relations}\). Finally, taking the limit as \(T \to 0\) and applying eq. \(\eqref{Mext}\), we have
\begin{equation}\label{Mextdgdepsilon}
    \lim_{T\to 0} \left(\frac{\partial M}{\partial \epsilon}\right)_{T,\vec{\mathcal{Q}}} = \lim_{T\to 0} \left(\frac{\partial G}{\partial \epsilon}\right)_{T,\vec{\mu}} \ .
\end{equation}
Now, considering the change in the $S$ at fixed $M, T,   \vec{\mathcal{Q}}$ we can verify that 
   \begin{equation}\label{dSdepsilon}
    -T\left (\frac{\partial S}{\partial \epsilon}\right )_{M,\vec{\mathcal{Q}}} = \left(\frac{\partial G}{\partial \epsilon}\right)_{T,\vec{\mu}} \ .
    \end{equation} 
From \eqref{Mextdgdepsilon} and \eqref{dSdepsilon}, the \eqref{Goon Penco relation} is verified. Moreover, the leading-order expansion of \eqref{Goon Penco relation} can be connected to the Weak Gravity Conjecture  by establishing proportional relationships between the higher derivatives of mass and entropy:
\begin{equation}
    \Delta M_{ext}(\Vec{\mathcal{Q}}) \approx -T_0(M,\Vec{\mathcal{Q}}) \Delta S(M,\Vec{\mathcal{Q}}) \bigg|_{M \approx M_{ext}^0(\Vec{\mathcal{Q}})} \ .
    \label{GPapp}
\end{equation}
This expansion is related to higher-derivative corrections, leading to \(\Delta S(M,\mathcal{Q}) \sim \Delta z > 0\) as demonstrated in \cite{Cheung:2018cwt}, where $\Delta z$ represents the shift in the charge-to-mass ratio. If our calculations align well with \eqref{Goon Penco relation}, it extends \eqref{GPapp} and supports the validity of the WGC.

\quad The action for anti-de Sitter (AdS) spacetime is typically written as the Einstein-Hilbert action with a negative cosmological constant \(\Lambda\), which reflects the curvature of the spacetime. The general form of the action is:
\begin{equation}\label{Master Action}
    \mathcal{I}_{\text{AdS}} = \frac{1}{2} \int d^4 x \sqrt{-g} \left( R - 2 \Lambda + \mathcal{I}_{\text{interaction}}\right) \ ,
\end{equation}
Here \(R\) is the Ricci scalar, \(\Lambda\) is the cosmological constant, and can be written in AdS radius $\ell$. This action describes the dynamics of pure gravity in AdS spacetime. Additional fields, such as gauge fields, can be included (as $\mathcal{I}_{\text{interaction}}$) to describe more complex systems. To get the Einstein-maxwell AdS black, the choice of $\mathcal{I}_{\text{interaction}}$ is $\mathscr{F}$. Here $\mathscr{F}$ is the Maxwell invariant and can be written as $F_{\mu \nu}F^{\mu \nu}$ where, \( F_{\mu\nu} \), is elegantly defined as \( F_{\mu\nu} = \partial_\mu A_\nu - \partial_\nu A_\mu \), where \( A_\mu \) represents the vector potential.

\subsection{Einstein-Power-Maxwell AdS Black Hole} \label{Sec:The power-law Maxwell electrodynamics}

The action for the  Einstein-Power-Maxwell AdS gravity can be obtained by using $\mathcal{I}_{\text{interaction}}$ as $(-\mathscr{F})^\gamma$ in \ref{Master Action}. The action for the Einstein-Power-Maxwell AdS gravity\cite{Hassane:2008} can be written as  
\begin{equation}\label{EPM Action}
\mathcal{I}_{\text{PM}} = \frac{1}{2}\, \int_{M}d^{4}x\sqrt{-g}\left[ R - 2\Lambda - \mathscr{F}^{\gamma}\right] \ ,  
\end{equation}
where $\Lambda = -3/\ell^2$, and $F_{\mu\nu} = \partial_\mu A_\nu - \partial_\nu A_\mu$ is the electromagnetic field tensor, with $ A_\mu$ being the vector potential, and $\gamma$ is the power-law exponent, which introduces non-linearity into the electromagnetic field. 

We perform a variation of the action with respect to the metric $g_{\mu \nu}$ to obtain the corresponding field equations. The field equations, derived through variations with respect to both the metric and the gauge field \( A_\mu \), take the following form respectively:
\begin{eqnarray}
    G_{\mu\nu} + \Lambda g_{\mu\nu} - \gamma (\mathscr{F})^{\gamma - 1} \left( F_{\mu\lambda} F_\nu{}^{\lambda} - \frac{1}{4}g_{\mu\nu} F_{\alpha\beta} F^{\alpha\beta} \right) &=& 0 \ , \nonumber \\
     \frac{1}{\sqrt{-g}}\nabla_\nu \left( \sqrt{-g} (\mathscr{F})^{\gamma-1} F^{\mu\nu} \right) &=& 0 \ .
\end{eqnarray}

\begin{figure}[ht]
    \begin{center}
        \includegraphics[scale=0.85]{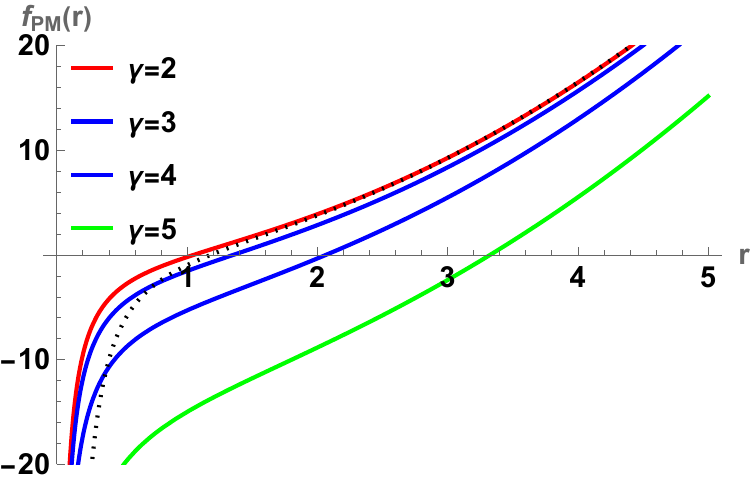}
    \end{center}
    \caption{Plot of metric function~\eqref{EPM metric} for different values of $\gamma$ here dotted line repersent the Maxwell case i.e., for $\gamma=1$.}
    \label{fig:fPM}
\end{figure}

Our goal is to find a spacetime geometry that is both static and spherically symmetric; we start with its line element expressed as
\begin{equation}\label{EPM metric}
    ds^{2} = -f_{\text{PM}}(r) dt^{2} + \frac{dr^{2}}{f_{\text{PM}}(r)} + r^2 \left( d\theta^2 + \sin^2\theta \, d\phi^2 \right) \ .
\end{equation}
By using the field equations derived from the variation of the bulk action using this metric, we can determine the metric function \( f(r) \) as (Plotted in Fig.\ref{fig:fPM})  
\begin{eqnarray}\label{EMP f(r)}
    f_{\text{PM}} (r) = 1+ \frac{r^2}{\ell^2}-\frac{2 M}{r}-\frac{(2 \gamma -1)^{2-2 \gamma} q^{2 \gamma}}{\left(2^{1-\gamma} (2 \gamma-3)^{1-2 \gamma}\right) r^{\frac{2}{2 \gamma-1}}} \ .
\end{eqnarray}

 The gauge potential one-form \( A \), and the electromagnetic field two-form \( F \) as
    \begin{eqnarray}
    A = - q r^{(2\gamma - 3)/(2\gamma - 1)} dt \;\;\;\;\;\;\;\;;\;\;\;\;\;\;\;  F = dA  \ , \nonumber
\end{eqnarray}
where the parameters \( M \) is the ADM mass, and \( q \) is an integration constant and related to the charge of the black hole as
\begin{equation}
    Q = 2^{\gamma-1}  \frac{(3-2\gamma)^{2\gamma-1}}{(2\gamma-1)^{2\gamma-2}} \; q^{2\gamma-1} \ .
\end{equation}
Also, \( \gamma \neq \frac{3}{2} \) represents the non-linearity parameter of the source, constrained by \( \gamma > \frac{1}{2} \) \cite{Hassane:2008}. 


\quad It has been demonstrated in \cite{Hassane:2008} that the metric defined by Eqs. (\ref{EPM metric}) and (\ref{EMP f(r)}) describes a black hole characterized by an event horizon \( r_{+} \) and a Cauchy horizon \( r_{-} \). The radius of the event horizon can be determined numerically by solving \( f_{\text{PM}}(r_{+}) = 0 \). The temperature of the black hole, derived from the surface gravity, is given by
\[
T = \frac{f_{\text{PM}}^{\prime}(r_{+})}{4\pi} = \frac{1}{4 \pi } \left[ \frac{2 r_+}{\ell^2}+\frac{2 M}{r_+^2}+2^{\gamma } \left(\frac{2}{1-2 \gamma }+1\right)^{2 \gamma -1} \frac{q^{2 \gamma }}{r_+^{\frac{2}{2 \gamma-1 }+1}}\right] \ .
\]
The mass can be expressed in terms of the horizon radius, which is
\begin{equation}
M =  -\frac{1}{2} r_+ \left(-\frac{r_+^2}{l^2}+2^{\gamma -1} \frac{(2 \gamma -3)^{2 \gamma -1}}{(2 \gamma -1)^{2 \gamma -2}} \frac{q^{2 \gamma }}{r_+^{\frac{2}{2 \gamma -1}}} - 1\right) \ .
\end{equation}

The electric potential \( \Phi \), measured at infinity relative to the horizon, and the black hole entropy \( S \), derived from the area law, are given by:
\[
\Phi =  \frac{q}{r_{+}^{(3-2\gamma)/(2\gamma-1)}} \;\;\;\;\;\;\;\;;\;\;\;\;\;\;\; S=\pi r_+^2,
\]
As previously considered \cite{Dolan:2011cqg1, Dolan:2011cqg2, Kubiznak:2012jhep, Dolan:2012jh}, we interpret the cosmological constant \( \Lambda \) as a thermodynamic pressure \( P \) and with the corresponding thermodynamic volume defined as \cite{Cvetic:2011prd}:

\[
P = -\frac{1}{8\pi}\Lambda = \frac{3}{8 \pi \ell^{2}} \;\;\;\;\;\;\;\;\;\;;\;\;\;\;\;\;\;\;\;\; V = \frac{4\pi r_{+}^{3}}{3}
\]
These quantities satisfy the Smarr relation:
\[
M = 2 \, T S + \frac{\Phi\, Q}{\gamma(2\gamma-1)}  -  V P \ ,
\]
which can be derived using scaling arguments \cite{Kastor:2009cqg, Gauntlett:1999cqg}. The first law of thermodynamics, i.e., $dM = TdS + \Phi dQ + VdP$, is also verified.

\subsection{Einstein-Power-Yang–Mills AdS Black Hole} \label{Sec:Einstein-Power-Yang–Mills AdS Black Hole}

The action for the  Einstein-Power-Yang-Mills AdS gravity can be obtained by using $\mathcal{I}_{\text{interaction}}$ as $(-\mathcal{F})^\xi$ in \ref{Master Action} where $\mathcal{F}$ is Yang-Mills invariant. The action for Einstein-power-Yang-Mills (EPYM) gravity with a cosmological constant $\Lambda$ is 
\begin{equation}\label{EPYM Action}
\mathcal{I}_{\text{PYM}} = \frac{1}{2} \int d^4x \sqrt{-g} \left[ R - 2 \Lambda - \mathcal{F}^\xi \right] \ ,
\end{equation}
and $\xi$ is a positive real parameter and $\mathcal{F}$ is the Yang-Mills invariant given by
\begin{eqnarray}
\mathcal{F} = \text{Tr}\left(F_{\lambda\sigma}^{(a)} F^{(a)\lambda\sigma}\right) \;\;\;\;\;\;\;\;\;\;;\;\;\;\;\;\;\;\;\;\; \text{Tr}(.) = \sum_{a=1}^{3} (.)  \nonumber 
\end{eqnarray}
The Yang-Mills field strength $F_{\lambda\sigma}^{(a)}$ is defined as
\begin{equation}
F_{\lambda\sigma}^{(a)} = \partial_\lambda A_\sigma^{(a)} - \partial_\sigma A_\lambda^{(a)} + \frac{1}{2\zeta} \Xi_{(b)(c)}^{(a)} A_\lambda^{(b)} A_\sigma^{(c)},
\label{YM}
\end{equation}
where $\Xi_{(b)(c)}^{(a)}$ are the structure constants of the Lie group with three parameters, $\zeta$ is the coupling constant, and $A_\mu^{(a)}$ represents the Yang-Mills potentials. The Yang-Mills (YM) field is defined as follows:
\begin{equation}
    F^{(a)} = dA^{(a)} + \frac{1}{2\sigma} \Xi^{(a)}_{(b)(c)} A^{(b)} \wedge A^{(c)} \ .
\end{equation}
Here, \( \Xi^{(a)}_{(b)(c)} \) represents the structure constants of the \( 3 \)-parameter Lie group \( G \), \( \zeta \) is a coupling constant, and \( A^{(a)} \) are the YM potentials corresponding to the SO(\(3\)) gauge group. The method for determining the components \( \Xi^{(a)}_{(b)(c)} \) has been detailed elsewhere \cite{Habib:2009grg}. It is important to note that the internal indices \( \{a, b, c, \ldots\} \) are invariant under covariant or contravariant transformations.

Varying the action with respect to the spacetime metric \( g_{\mu\nu} \) leads to the field equations:
\[
G_{\mu\nu} +  \Lambda g_{\mu\nu} = -\frac{1}{2} \left( \delta_{\nu}^\mu \mathcal{F}^\xi - 4\xi \, \text{Tr} \left[ F^{(a)}_{\nu\lambda} F^{(a)\mu\lambda} \right] \mathcal{F}^{\xi-1} \right) \ ,
\]
here, \( G_{\mu\nu} \) denotes the Einstein tensor. Variation with respect to the gauge potentials \( A^{(a)} \) gives rise to the YM equations:
\[
d \left( ^\ast F^{(a)} \mathcal{F}^{\xi-1} \right) + \frac{1}{\zeta} \Xi^{(a)}_{(b)(c)} \mathcal{F}^{\xi-1} A^{(b)} \wedge {^\ast F^{(c)}} = 0
\]
where \( \ast \) denotes the Hodge dual operator.

The metric function $f_{\text{PYM}}(r)$ for a 4-dimensional EPYM black hole with a negative cosmological constant is
\begin{equation}\label{EPYM}
f_{\text{PYM}}(r) = 1 - \frac{2M}{r} + \frac{r^2}{\ell^2} + \frac{(2Q^2)^\xi}{2(4\xi-3) r^{4\xi-2}} \ .
\end{equation}

\begin{figure}[ht]
    \begin{center}
        \includegraphics[scale=0.85]{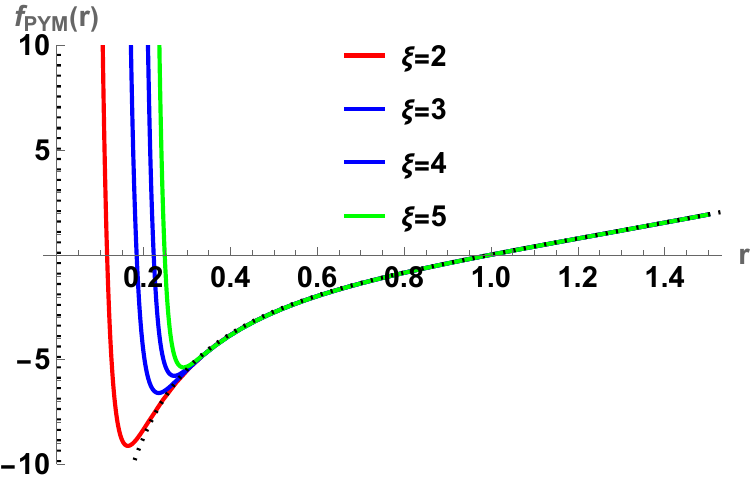}
    \end{center}
    \caption{Plot of metric function~\eqref{EPYM} for different values of $\xi$ here dotted line repersent the Yang-Mills case i.e., for $\xi=1$.}
    \label{fig:fPYM}
\end{figure}
Here, $M$ is related to the black hole mass, and $Q$ denotes the charge parameter associated with the Yang-Mills fields. Fig.\ref{fig:fPYM}, show the shift in the horizon with parameter $\xi$. To ensure the Weak Energy Condition (WEC) for the Power-Yang-Mills term, $\xi$ must be positive \cite{Mazharimousavi:2009mb}. The parameter $\xi$ is constrained by the energy and causality conditions discussed in \cite{Mazharimousavi:2009mb}, specifically $\frac{3}{4} \leq \xi < 2$. For $\xi = 1$, the solutions reduce to the Einstein-Yang-Mills black holes in four dimensions \cite{HabibMazharimousavi:2007fst, Mazharimousavi:2008ap}.

The position of the event horizon, denoted by \( r_h \), is identified as the largest positive real root of the equation \( f_{\text{PYM}}(r_h) = 0 \) derived from Eq. (\ref{EPYM}). The mass of the black hole in terms of horizon radius is 
\begin{equation*}
    M =  \frac{1}{4} r_h \left(\frac{2 r_h^2}{\ell^2}+\frac{2^{\xi } r_h^{2-4 \xi }}{4 \xi -3} Q^{2 \xi } + 2\right) \ .
\end{equation*}
The temperature of the black hole, as determined by its surface gravity, is expressed as
\begin{equation*}
    T = \frac{1}{4 \pi } \left[\frac{2 r_h}{l^2}+\frac{2 M}{r_h^2}+\frac{2^{\xi } (1-2 \xi )  r_h^{1-4 \xi }}{4 \xi -3} Q^{2 \xi } \right] \ . 
\end{equation*}
It's already discussed in \cite{Zhang:2015grg}, the YM potential and the entropy of the black hole is 
\begin{equation*}
    \Phi_Q = \frac{ \xi (2Q^2)^\xi}{2(4\xi-3)Q} r_h^{3-4\xi} \;\;\;\;\;\;\;\;\;\;;\;\;\;\;\;\;\;\;\;\; S=\pi r_h^2 \ .
\end{equation*}

 The Smarr relation for an Einstein-Power-Yang-Mills (EPYM) black hole in the extended phase space is derived by utilizing all the aforementioned quantities and treating the mass \( M \) as the enthalpy of the black hole \cite{Kastor:2009cqy}:
\[
M = 2 TS + \frac{2\xi-1}{\xi} \Phi_Q Q - 2 VP
\]
However, using the expressions for the thermodynamic quantities must satisfy the first law of thermodynamics, i.e., $dM = TdS + \Phi_Q dQ + VdP$.

\subsection{Einstein-Maxwell-Power-Yang-Mills AdS black holes} \label{Sec:The power-law Yang Mills electrodynamics}

The action for a Maxwell-power-Yang-Mills theory in an anti-de Sitter (AdS) spacetime is given by:
\begin{equation}\label{EMPYM Action}
\mathcal{I}_{\rm MPYM} = \frac{1}{2} \int d^4x \sqrt{-g} \left[ \left(R - 2\Lambda \right) - F_{\mu\nu} F^{\mu\nu} - \mathcal{F}^\xi \right] \ ,
\end{equation}
Where $R$ is the Ricci scalar, $\Lambda$ is the cosmological constant, $F_{\mu\nu}$ is the Maxwell field strength tensor, $\mathcal{F}_{\mu\nu}$ is the Yang-Mills field strength tensor and $\xi$ is the power-law exponents for the Yang-Mills fields, respectively. 

\begin{figure}[ht]
    \begin{center}
        \includegraphics[scale=0.85]{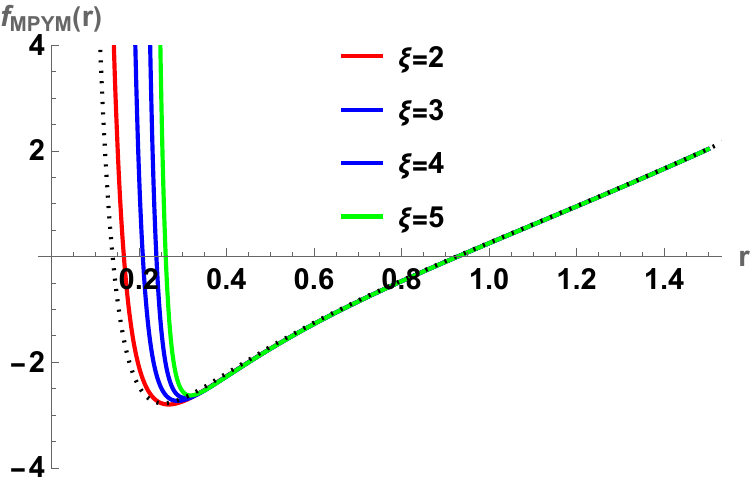}
    \end{center}
    \caption{Plot of metric function~\eqref{EMPYM} for different values of $\xi$.}
    \label{fig:fMPYM}
\end{figure}

The metric function $f(r)$ for a $4$-dimensional spherically symmetric line element is 
\begin{equation}\label{EMPYM}
    f_{\text{MPYM}}(r) = 1 - \frac{2M}{r} + \frac{r^2}{\ell^2} + \frac{Q_M}{r^2} + \frac{(2 Q_{YM})^\xi}{2(4\xi-3)r^{4\xi-2}} \ ,
\end{equation}
where parameter \( M \) represents the mass of the black hole, parameter $\xi \neq 3/4$ while \( Q_M \) and \( Q_{YM} \) correspond to the charges associated with the Maxwell field and the Yang-Mills field, respectively and plotted in Fig.\ref{fig:fMPYM}.

The location of the event horizon, denoted as \( r_h \), is determined as the largest positive real root of the equation \( f_{\text{MPYM}}(r)|_{r=r_h} = 0 \), derived from Eq. (\ref{EMPYM}). The mass of the black hole in terms of the horizon radius is 
\[
M =  \frac{1 }{4 r_h} \left[ 2 Q_M^2+\frac{2 r_h^4}{\ell^2}+\frac{2^{\xi } \left(Q_{YM}\right)^{2\xi } r_h^{4-4 \xi }}{4 \xi -3}+2 r_h^2 \right] \ .
\]
The temperature of the black hole, calculated from its surface gravity, is expressed as:
\[
T = \frac{1}{4 \pi } \left[ -\frac{2 Q_M^2}{r_h^3}+\frac{2 r_h}{\ell^2}+\frac{2 M}{r_h^2}+\frac{2^{\xi } (1-2 \xi ) \left(Q_{YM}\right)^{2 \xi } r_h^{1-4 \xi }}{4 \xi -3} \right] \ . 
\]
The Yang-Mills potential and the Maxwell potential are
\[
\Phi_{Q_M}= \frac{Q_M}{r_h}   \;\;\;\;\;\;\;\;\;\; \text{and} \;\;\;\;\;\;\;\;\;  \Phi_{Q_{YM}} = \frac{\xi (2Q^2)^\xi}{2(4\xi - 3)Q} r_h^{3 - 4\xi} \ .
\]
The Smarr relation for an Einstein-Power-Yang-Mills (EPYM) black hole in the extended phase space is derived by incorporating the above thermodynamic quantities and treating the mass \( M \) as the black hole's enthalpy \cite{Kastor:2009cqy}:
\[
M = 2 TS +\Phi_{Q_M}Q_M + \frac{2\xi-1}{\xi } \Phi_{Q_{YM}} Q_{YM} - 2VP \ .
\]
Furthermore, the first law of thermodynamics must be satisfied by these expressions for the thermodynamic quantities as $dM = TdS + \Phi_{Q_M} dQ_M +\Phi_{Q_{YM}} dQ_{YM} + VdP$.

\section{ Extremality Relations}\label{Sec: Extrmality Relations}

This section will introduce correction terms as specified in equation \eqref{Delta I}. By modifying the action, the metric will also undergo corrections, affecting the thermodynamic quantities. We will derive these thermodynamic quantities in terms of the perturbation parameter and check whether equation \eqref{Goon Penco relation} holds true.

\subsection{Universal Relation on the Power-Maxwell AdS black hole}
In this subsection, we will add the correction term in the Power-Maxwell AdS black hole action as defined in equation \eqref{EPM Action}. By incorporating this correction, only the $dt$ and $dr$ terms are affected. Consequently, the metric is corrected as
\begin{equation*}
   ds^2_{\text{Perturbed}} = -\frac{r^2}{\ell^2} dt^2 - \frac{r^2  }{\ell^2 f_{\text{PM}}(r)} dr^2 \ .
\end{equation*}
We equate the metric function $f(r)$ obtained from the total action by zero to compute the perturbed mass. In terms of entropy $S$, the perturbed mass\footnote{This mass is computed at $T=0$ limit so, this mass is same as the $M_{\text{ext}}$.} is 
\begin{eqnarray}\label{EPM Mass perturbed}
  M_{\text{ext}}( \Vec{\mathcal{Q}}, \epsilon ) &=& \frac{\sqrt{S}}{2 \sqrt{\pi }} \left( 1 + \frac{S(\epsilon +1)}{\pi  \ell^2}-2^{\gamma-1} \pi ^{\frac{1}{2 \gamma-1}} \frac{(2 \gamma-3)^{2 \gamma-1}}{ (2 \gamma-1)^{2 \gamma-2}} \frac{q^{2 \gamma}}{S^{\frac{1}{2 \gamma-1}}}\right) \ .
\end{eqnarray}

\begin{figure}[ht]
    \begin{center}
        \includegraphics[scale=0.65]{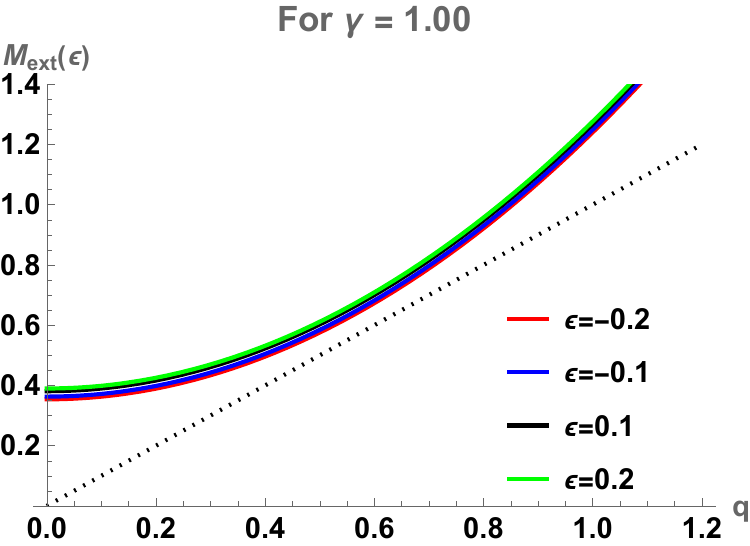}
        \hspace{0.2cm}
        \includegraphics[scale=0.65]{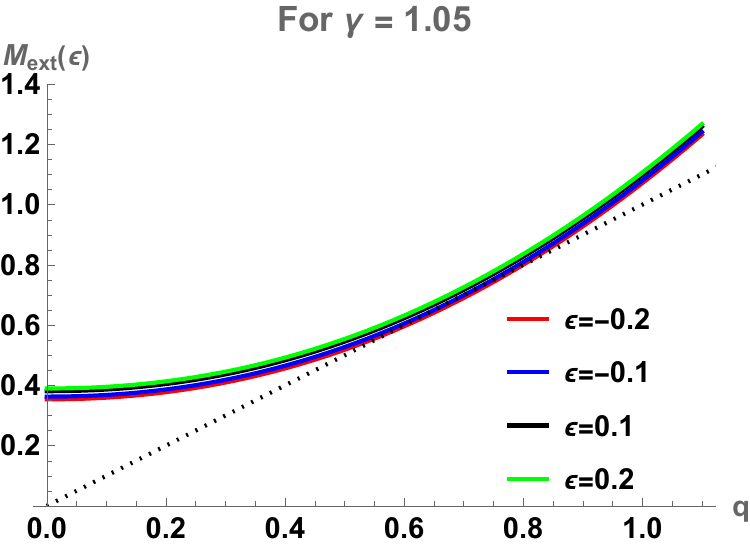}
    \end{center}
    \caption{Plot of Eq.~\eqref{EPM Mass perturbed} vs $q$ for different perturbation parameters to observe the WGC-like behaviour. It can be observed that negative correction results in a mass decrease.}
    \label{fig:PM}
\end{figure}

Here $\Vec{\mathcal{Q}}$ denotes all the fluctuating quantities. To find the perturbed temperature, we calculate the surface gravity, and the expression for perturbed temperature in terms of entropy is 
\begin{eqnarray}\label{EPM temperature perturbed}
   T\;(\Vec{\mathcal{Q}}, \epsilon) &=& \frac{\epsilon +1}{8 \pi ^{3/2} \ell^2 \sqrt{S}} \left[\frac{\pi  \ell^2 }{\epsilon +1} \left(2-2^\gamma \pi ^{\frac{1}{2 \gamma-1}} \frac{(2 \gamma-3)^{2 \gamma}}{ (2 \gamma-1)^{2 \gamma-1}} \frac{q^{2 \gamma}}{ S^{\frac{1}{2 \gamma-1}}}\right)+6 S\right] \ . 
\end{eqnarray}
Inverting Eq.\eqref{EPM Mass perturbed}, the value of perturbation parameter $\epsilon$ is 
\begin{equation}\label{EPM Epsilon}
    \epsilon =  -\frac{ \ell^2 }{2 S^{3/2}} \left[ \frac{2 S^{3/2}}{\ell^2}-2^\gamma \pi ^{\frac{1}{2 \gamma-1}+1} \frac{(2 \gamma-3)^{2 \gamma-1}}{ (2 \gamma-1)^{2 \gamma-2}} \frac{q^{2 \gamma}} {S^{\frac{1}{2 \gamma-1}-\frac{1}{2}}}+2 \pi  \sqrt{S}-4 M \pi ^{3/2}\right] \ .
\end{equation}
Now, by taking its derivative w.r.t. $S$ and its inverse, we can easily calculate the R.H.S of equation \eqref{Goon Penco relation}. The expression for R.H.S is
\begin{equation}\label{Tds for Power maxwell}
-T\left(\frac{\partial S}{\partial \epsilon}\right)_{M,Q} =\frac{S^2  \left[\pi  \ell^2 \left(2-2^\gamma \pi ^{\frac{1}{2 \gamma-1}} \frac{(2 \gamma-3)^{2 \gamma}} {(2 \gamma-1)^{2 \gamma-1}} \frac{q^{2 \gamma}} {S^{\frac{1}{2 \gamma-1}}}\right)+6 S (\epsilon +1)\right]}{8 \pi ^{5/2} \ell^4 \left[-3 \sqrt{\pi } M-2^\gamma \pi ^{\frac{1}{2 \gamma-1}} \left(\frac{2}{1-2 \gamma}+1\right)^{2 \gamma-1} \gamma \frac{q^{2 \gamma}} {S^{\frac{1}{2 \gamma-1}}+\frac{1}{2}}+\sqrt{S}\right]}
\end{equation}
Now, plugging the value of perturbation parameter epsilon from equation \eqref{EPM Epsilon}, we can easily see that the universal relation is satisfied, i.e., 
\begin{equation}
     -T \left( \frac{\partial S}{\partial\epsilon} \right)_{M,\Vec{\mathcal{Q}}} = \left(\frac{\partial M_{\text{ext}}}{\partial \epsilon} \right)_{\vec{\mathcal{Q}}} = \frac{r_+^3}{2\ell^2}  \ .
\end{equation}
The structure of equation \eqref{Goon Penco relation} is rigorously upheld within the framework of power-law Maxwell black holes. 

Using the first law of thermodynamics, we can also have another form of extremality relation. For this now, we will assume the pressure, i.e., the cosmological constant is not constant, then we have the relation of perturbation parameter in terms of pressure $P$ as
\begin{equation*}
   \epsilon(P) = -\frac{1 }{16 P S^{3/2}} \left[16 P S^{3/2}-3\ 2^\gamma \pi ^{\frac{1}{2 \gamma-1}} \frac{(2 \gamma-3)^{2 \gamma-1}} {(2 \gamma-1)^{2 \gamma-2}} \frac{q^{2 \gamma}} {S^{\frac{1}{2 \gamma-1}-\frac{1}{2}}}+6 \sqrt{S} - 12 M \sqrt{\pi} \right] \ .
\end{equation*}
Again taking its derivative w.r.t $P$ and using the relation for $V$\footnote{Here $V=\frac{4}{3}\pi r_+^3$} we have
\begin{equation*}
    -V\left(\frac{\partial P}{\partial \epsilon}\right) = \frac{64 P^2 S^3}{3 \sqrt{\pi } \left(-12 \sqrt{\pi } M-3\ 2^\gamma \pi ^{\frac{1}{2 \gamma-1}} \frac{(2 \gamma-3)^{2 \gamma-1}} {(2 \gamma-1)^{2 \gamma-2}} \frac{q^{2 \gamma}} {S^{\frac{1}{2 \gamma-1}-\frac{1}{2}}}+6 \sqrt{S}\right)} \ .
\end{equation*}
Again by plugging the form of mass $M$, the expression satisfies another form of extremality relation, i.e.,
\begin{equation*}
    - V\left(\frac{\partial P}{\partial \epsilon}\right) = \left(\frac{\partial M_{\text{ext}}}{\partial \epsilon} \right)_{\vec{\mathcal{Q}}} = \frac{r_+^3}{2\ell^2}  \ .
\end{equation*}
So, other forms of Goon and Penco are also satisfied in the case of Einstein-power Maxwell black holes.

\subsection{Universal Relation on the Power-Yang-Mills AdS black hole}

In this subsection, we introduce a correction term to the action of the Power-Yang-Mills AdS black hole, as specified in equation \eqref{EPYM Action}. The incorporation of this correction modifies only the \( dt \) and \( dr \) components, leading to a corrected metric as 
\[
ds^2_{\text{Perturbed}} = -\frac{r^2}{\ell^2} dt^2 - \frac{r^2}{\ell^2 f_{\text{PYM}}(r)} dr^2 \ .
\]
To compute the perturbed mass, we set the metric function \( f_{\text{PYM}}(r) \), derived from the total action, to zero. Expressed in terms of entropy \( S \), the perturbed mass is given by:

\begin{equation}\label{EPYM Mass perturbed}
   M_{\text{ext}}( \Vec{\mathcal{Q}}, \epsilon ) = \frac{1}{4 \pi ^{3/2}} \left[\frac{2 \sqrt{S} \left(\pi  l^2+S \epsilon +S\right)}{l^2}+\frac{2^{\xi } \pi ^{2 \xi } \left(Q^2\right)^{\xi } S^{\frac{3}{2}-2 \xi }}{4 \xi -3}\right] \ ,
\end{equation}
here, the vector $\Vec{\mathcal{Q}}$ represents all the fluctuating parameters.

\begin{figure}[ht]
    \begin{center}
        \includegraphics[scale=0.65]{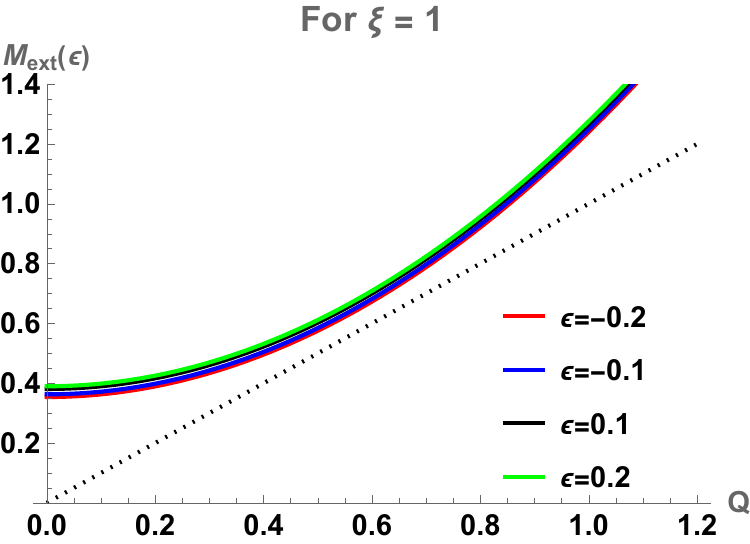}
        \hspace{0.2cm}
        \includegraphics[scale=0.65]{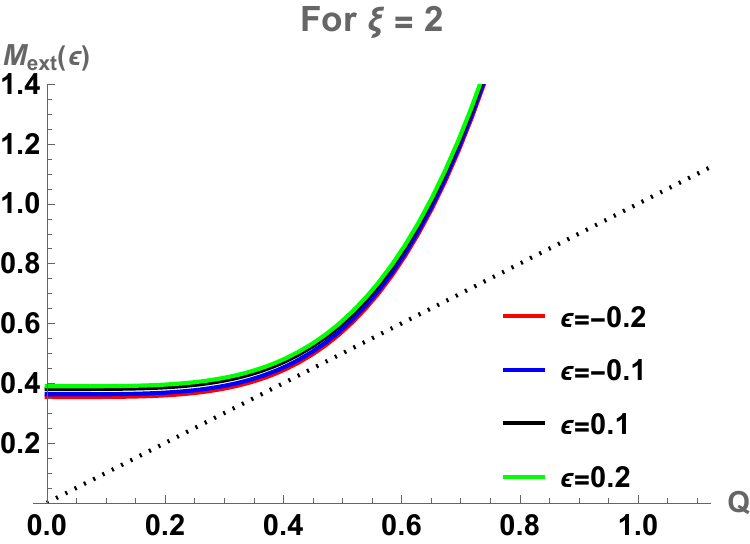}
    \end{center}
    \caption{Plot of Eq.~\eqref{EPYM Mass perturbed} vs $q$ for different perturbation parameters to observe the WGC-like behaviour. It can be observed that negative correction results in a mass decrease.}
    \label{fig:PYM}
\end{figure}

To derive the perturbed temperature, the surface gravity is calculated, leading to the expression for the perturbed temperature in terms of entropy:
\begin{equation}\label{EPYM temperature perturbed}
T(\Vec{\mathcal{Q}}, \epsilon) = \frac{S^{-2 \xi -\frac{1}{2}} }{8 \pi ^{3/2} \ell^2}  \left[2 (\epsilon +1) \left(\pi  \ell^2+3 S\right) S^{2 \xi }-2^{\xi } \pi ^{2 \xi } \ell^2 S \left(Q^2\right)^{\xi }\right] \ .
\end{equation}
By inverting equation \eqref{EPYM Mass perturbed}, the perturbation parameter $\epsilon$ is determined as follows:
\begin{equation}\label{EPYM Epsilon}
\epsilon = \ell^2 \left(\frac{2 \pi ^{3/2} M}{S^{3/2}}+\frac{2^{\xi -1} \pi ^{2 \xi } \left(Q^2\right)^{\xi } S^{-2 \xi }}{3-4 \xi }-\frac{\pi }{S}\right)-1 \ .
\end{equation}
Taking the derivative of this expression with respect to $S$ and then inverting it allows for a straightforward calculation of the right-hand side (R.H.S) of equation \eqref{Goon Penco relation}. The expression for the R.H.S is:
\begin{equation}
-T\left(\frac{\partial S}{\partial \epsilon}\right)_{M,Q} = \frac{(4 \xi -3) S^2 \left(2 (\epsilon +1) \left(\pi  l^2+3 S\right) S^{2 \xi }-2^{\xi } \pi ^{2 \xi } l^2 S \left(Q^2\right)^{\xi }\right)}{8 \pi ^{3/2} l^4 \left(-3 \pi ^{3/2} (4 \xi -3) M S^{2 \xi }+2^{\xi } \pi ^{2 \xi } \xi  S^{3/2} \left(Q^2\right)^{\xi }+\pi  (4 \xi -3) S^{2 \xi +\frac{1}{2}}\right)} \ .
\end{equation}
Substituting the value of the perturbation parameter $\epsilon$ from equation \eqref{EPYM Epsilon}, it becomes evident that the universal relation holds true:
\begin{equation}\label{Tds for Power YM}
-T \left(\frac{\partial S}{\partial \epsilon}\right)_{M, \Vec{\mathcal{Q}}} = \left(\frac{\partial M_{\text{ext}}}{\partial \epsilon}\right)_{\Vec{\mathcal{Q}}} = \frac{1}{2\ell^2} \left(\frac{S}{\pi}\right)^{3/2} \ .
\end{equation}
This confirms that the structure of equation \eqref{Goon Penco relation} is consistently maintained within the framework of power-law-Yang-Mills black holes. This consistency is observed regardless of the variations in the power-law exponent, indicating that the value remains unaffected by such changes.

Furthermore, another form of the extremality relation can be derived by invoking the first law of thermodynamics. If we assume that the pressure, which is associated with the cosmological constant, is variable, the perturbation parameter $\epsilon$ can be expressed in terms of pressure $P$ as follows:
\begin{equation}
\epsilon(P) = \frac{3}{16 P} \left[\frac{4 \sqrt{\pi } M}{S^{3/2}}+\frac{2^{\xi } \pi ^{2 \xi -1} \left(Q^2\right)^{\xi } S^{-2 \xi }}{3-4 \xi }-\frac{2}{S}\right] - 1 \ .
\end{equation}
Taking the derivative of this expression with respect to $P$ and utilizing the relation for $V$, we obtain:
\begin{equation}
-V\left(\frac{\partial P}{\partial \epsilon}\right) = \frac{64 P^2 S^{3/2}}{9 \sqrt{\pi } } \left[-\frac{4 \sqrt{\pi } M}{S^{3/2}}+\frac{2^{\xi } \pi ^{2 \xi -1} \left(Q^2\right)^{\xi } S^{-2 \xi }}{4 \xi -3}+\frac{2}{S}\right]^{-1} \ .
\end{equation}
Again, by substituting the mass $M$ into this expression, it is evident that this form of the extremality relation is satisfied:
\begin{equation}
-V\left(\frac{\partial P}{\partial \epsilon}\right) = \left(\frac{\partial M_{\text{ext}}}{\partial \epsilon}\right)_{\Vec{\mathcal{Q}}} = \frac{1}{2\ell^2} \left(\frac{S}{\pi}\right)^{3/2} \ .
\end{equation}
Therefore, different forms of the Goon-Penco relation are also valid in Einstein-power-Yang-Mills black holes.

Before closing this subsection, one more form of  the Goon-Penco relation as 
\begin{equation*}
    -\Phi_Q \left(\frac{\partial Q}{\partial \epsilon}\right) =  \left(\frac{\partial M_{\text{ext}}}{\partial \epsilon}\right)_{\Vec{\mathcal{Q}}} = \frac{1}{2\ell^2} \left(\frac{S}{\pi}\right)^{3/2} \ .
\end{equation*}

\subsection{Universal Relation on the Maxwell-Power-Yang-Mills AdS black hole} 

In this subsection, we introduce a perturbative correction to the action of the Power-Yang-Mills AdS black hole, as detailed in equation \eqref{EMPYM Action}. This correction impacts only the \( dt \) and \( dr \) components of the metric, resulting in the following perturbed line element:
\[
ds^2_{\text{Perturbed}} = -\frac{r^2}{\ell^2} dt^2 + \frac{r^2}{\ell^2 f_{\text{MPYM}}(r)} dr^2 \ .
\]
To calculate the perturbed mass, we set the metric function \( f_{\text{MPYM}}(r) \), derived from the complete action, to zero. The resulting expression for the perturbed mass, in terms of the entropy \( S \), is given by:
\begin{equation}\label{EMPYM Mass perturbed}
   M_{\text{ext}}( \Vec{\mathcal{Q}}, \epsilon ) = \frac{\sqrt{S} }{2 \sqrt{\pi }}  \left[\frac{1 }{2 \pi } S \left(\frac{2 (\epsilon +1)}{\ell^2}+\frac{2^{\xi } \pi ^{2 \xi } S^{-2 \xi } \left(Q_{\text{YM}}^2\right){}^{\xi }}{4 \xi -3}\right)+\frac{\pi  Q_M^2}{S}+1\right] \ ,
\end{equation}
where \( \Vec{\mathcal{Q}} \) encapsulates all fluctuating parameters. The perturbed temperature is then obtained by evaluating the surface gravity, yielding the following expression in terms of the entropy:
\begin{equation}\label{EMPYM temperature perturbed}
T(\Vec{\mathcal{Q}}, \epsilon) =\frac{S^{-2 \xi -\frac{3}{2}} \left(2 S^{2 \xi } \left(S \left(\pi  \ell^2+3 S (\epsilon +1)\right)-\pi ^2 \ell^2 Q_M^2\right)-2^{\xi } \pi ^{2 \xi } \ell^2 S^2 \left(Q_{\text{YM}}^2\right){}^{\xi }\right)}{8 \pi ^{3/2} \ell^2} \ .
\end{equation}
The perturbation parameter \( \epsilon \) is derived by inverting equation \eqref{EMPYM Mass perturbed} as follows:
\begin{equation}\label{EMPYM Epsilon}
\epsilon = \frac{1}{2} \ell^2 \left[\frac{2^{\xi } \pi ^{2 \xi } S^{-2 \xi } \left(Q_{\text{YM}}^2\right){}^{\xi }}{3-4 \xi }-\frac{2 \pi  \left(\pi  Q_M^2-2 \sqrt{\pi } M \sqrt{S}+S\right)}{S^2}\right]-1 \ .
\end{equation}
By differentiating this expression with respect to \( S \) and subsequently inverting, the right-hand side (R.H.S) of equation \eqref{Goon Penco relation} can be computed. The expression for the R.H.S is:
\begin{equation}
-T\left(\frac{\partial S}{\partial \epsilon}\right)_{M,Q} = \frac{(\epsilon +1) S^{\frac{3}{2}-2 \xi } \left(\frac{l^2 \left(-2 \pi  S^{2 \xi } \left(\pi  Q_M^2-S\right)-2^{\xi } \pi ^{2 \xi } S^2 \left(Q_{\text{YM}}^2\right){}^{\xi }\right)}{\epsilon +1}+6 S^{2 \xi +2}\right)}{8 \pi ^{3/2} l^4 \left(\pi  \left(2 \pi  Q_M^2-3 \sqrt{\pi } M \sqrt{S}+S\right)+\frac{2^{\xi } \pi ^{2 \xi } \xi  S^{2-2 \xi } \left(Q_{\text{YM}}^2\right){}^{\xi }}{4 \xi -3}\right)} \ .
\end{equation}
Substituting the expression for the perturbation parameter \( \epsilon \) from equation \eqref{EMPYM Epsilon} confirms the universal relation:
\begin{equation}\label{Tds for Power YM}
-T \left(\frac{\partial S}{\partial \epsilon}\right)_{M, \Vec{\mathcal{Q}}} = \left(\frac{\partial M_{\text{ext}}}{\partial \epsilon}\right)_{\Vec{\mathcal{Q}}} = \frac{1}{2\ell^2} \left(\frac{S}{\pi}\right)^{3/2} \ .
\end{equation}
This demonstrates that the structure of equation \eqref{Goon Penco relation} remains intact within the framework of Maxwell power-Yang-Mills black holes, irrespective of variations in the power-law exponent, highlighting the robustness of the relation against such changes.

Furthermore, an alternative form of the extremality relation can be derived by invoking the first law of thermodynamics. If we allow the pressure, related to the cosmological constant, to vary, the perturbation parameter \( \epsilon \) can be expressed as a function of the pressure \( P \) as follows:
\begin{equation}
\epsilon(P) = \frac{3} {16 \pi  P} \left[\frac{2^{\xi } \pi ^{2 \xi } \left(Q_{YM}^2\right)^{\xi } S^{-2 \xi }}{3-4 \xi }-\frac{2 \pi  \left(\pi  Q_M^2-2 \sqrt{\pi } M \sqrt{S}+S\right)}{S^2}\right]-1 \ .
\end{equation}
Taking the derivative with respect to \( P \) and applying the relation for \( V \), we obtain:
\begin{equation}
-V\left(\frac{\partial P}{\partial \epsilon}\right) = -\frac{64 \sqrt{\pi } P^2 S^{3/2}}{9 \left(\frac{2^{\xi } \pi ^{2 \xi } S^{-2 \xi } \left(Q_{\text{YM}}^2\right){}^{\xi }}{3-4 \xi }-\frac{2 \pi  \left(\pi  Q_M^2-2 \sqrt{\pi } M \sqrt{S}+S\right)}{S^2}\right)} \ .
\end{equation}
Once again, substituting the mass \( M \) into this equation shows that this form of the extremality relation is satisfied:
\begin{equation}
-V\left(\frac{\partial P}{\partial \epsilon}\right) = \left(\frac{\partial M_{\text{ext}}}{\partial \epsilon}\right)_{\Vec{\mathcal{Q}}} = \frac{1}{2\ell^2} \left(\frac{S}{\pi}\right)^{3/2} \ .
\end{equation}
Thus, multiple formulations of the Goon-Penco relation are applicable in the context of Einstein-Power-Yang-Mills black holes.

Before concluding this subsection, we present one additional form of the Goon-Penco relation:
\begin{equation*}
    -\Phi_{Q_M} \left(\frac{\partial Q_M}{\partial \epsilon}\right) =  \left(\frac{\partial M_{\text{ext}}}{\partial \epsilon}\right)_{\Vec{\mathcal{Q}}} = \frac{1}{2\ell^2} \left(\frac{S}{\pi}\right)^{3/2} \ ,
\end{equation*}
and the same from Yang-Mills charge and potential as well
\begin{equation*}
    -\Phi_{Q_{YM}} \left(\frac{\partial Q_{YM}}{\partial \epsilon}\right) =  \left(\frac{\partial M_{\text{ext}}}{\partial \epsilon}\right)_{\Vec{\mathcal{Q}}} = \frac{1}{2\ell^2} \left(\frac{S}{\pi}\right)^{3/2} \ ,
\end{equation*}

\section{Discussion}
\label{Sec: Discussion}

This study explores the thermodynamic relationships associated with modified thermodynamic quantities in the context of perturbative corrections to the actions of Maxwell-Power and Power-Yang-Mills black holes. The perturbative correction is introduced by adding a term related to the cosmological constant into the action. We carefully calculate the extremality relations that connect mass, pressure, entropy, and charges. Notably, we find that these extremality relations are equivalent, which can be explained by the first law of thermodynamics. In our analysis, we treat the cosmological constant as a variable linked to pressure, identifying its conjugate variable as the thermodynamic volume. This adjustment allows us to maintain the extremality relationship between mass and pressure, even with the rescaled constant included in the action. Our results show that the shifted mass bounds increase with the correction parameter $\epsilon$ for both types of black holes studied. This suggests that such corrections could enable the black hole to meet the Weak Gravity Conjecture conditions, similar to behaviors observed in other charged black holes. The relationship uncovered in this study demonstrates a greater universality than previously recognized. The proportional connection among the adjusted mass, temperature, and entropy offers valuable insights into the Weak Gravity Conjecture. Consequently, these findings contribute to a more profound understanding of quantum gravity. Additionally, they are anticipated to inspire and motivate further research endeavors in this domain

\quad Figures \ref{fig:PM}, and \ref{fig:PYM} depict the variation of modified black hole mass as a function of charge under different power for perturbation parameter corrections \(\epsilon\). In these plots, specific parameters are held constant, enabling a comparative analysis of black hole mass modifications due to small positive and negative corrections. Initially, the unmodified black hole state is characterized by a mass-to-charge ratio of unity. However, upon introducing constant corrections, the modified black hole exhibits a mass-to-charge ratio greater than one. We consider different correction values, specifically \(\epsilon = -0.2, -0.1, 0.1, 0.2\), and observe distinct variations in the black hole mass for each case. Positive corrections lead to an increase in the black hole mass, whereas negative corrections result in a mass reduction. This effect is inversely reflected in the entropy calculations, where a positive correction decreases entropy. Notably, when a small negative correction parameter is introduced, the black hole mass decreases toward unity, effectively increasing the charge-to-mass ratio (or equivalently reducing the mass-to-charge ratio). This behavior is consistent with the predictions of the Weak Gravity Conjecture (WGC), suggesting a WGC-like effect. The WGC, proposed to ensure the absence of stable, super-extremal black holes, serves as a guiding principle in delineating the boundary between consistent and inconsistent low-energy effective field theories. Our findings reinforce this conjecture by illustrating that modified gauge field dynamics in Power-Maxwell and Power-Yang–Mills frameworks still conform to WGC expectations. This provides further evidence supporting Vafa’s Swampland conjectures , which suggest that viable low-energy theories must respect quantum gravity constraints.

\quad In conclusion, our comprehensive analysis, particularly regarding the black holes examined, underscores the universal nature of the relationships. By rigorously investigating the proportional relationship between the corrected mass and entropy, we identify a significant pathway for enhancing our understanding of the weak gravity conjecture and help in the understanding of how modified electrodynamics and non-Abelian gauge theories interact with gravitational systems within the Swampland framework. By validating WGC-like behavior in the presence of generalized gauge fields, we offer new insights into the consistency of these theories in the quantum gravity landscape. The Weak Gravity Conjecture is crucial for Vafa’s Swampland program, ensuring that consistent low-energy theories comply with quantum gravity constraints. It prevents stable extremal black holes and constrains the landscape of viable, effective field theories, reinforcing Vafa’s conjecture that only certain theories reside in the quantum gravity landscape.
.

\section*{Acknowledgment}
I am grateful to Saeed Noori Gashti for discussions on related topics.






\end{document}